\documentclass{article}

\usepackage[preprint]{neurips_2026}

\usepackage[utf8]{inputenc}
\usepackage[T1]{fontenc}
\usepackage{hyperref}
\usepackage{url}
\usepackage{booktabs}
\usepackage{amsfonts}
\usepackage{amsmath}
\usepackage{microtype}
\usepackage{xcolor}
\usepackage{graphicx}
\usepackage{caption}
\usepackage{subcaption}
\usepackage{xspace}
\usepackage{pifont}
\usepackage{multirow}
\usepackage{enumitem}
\usepackage{tikz}
\usepackage{comment}
\usetikzlibrary{positioning,arrows.meta,fit,backgrounds,calc}

\makeatletter
\g@addto@macro{\UrlBreaks}{\UrlOrds}
\makeatother
\newcommand{\ourmethod}{\textsc{Plume}\xspace}
\newcommand{\ourtwo}{\textsc{Plume-Deep}\xspace}
\newcommand{\plumemamba}{\textsc{Plume-Mamba}\xspace}
\newcommand{\netssm}{\textsc{NetSSM}\xspace}
\newcommand{\commentA}[1]{}
\newcommand{\commentB}[1]{}
\newcommand{\commentC}[1]{}

\title{Protocol-Aware Tokenization and Architecture Co-Design for Wireless Packet Foundation Models}

\author{%
  Swadhin Pradhan \\
  Cisco Systems \\
  \texttt{swapradh@cisco.com}
  \And
  Shazal Irshad \\
  Cisco Systems \\
  \texttt{sirshad@cisco.com}
  \And
  Jerome Henry \\
  Cisco Systems \\
  \texttt{jerhenry@cisco.com}
}

\begin{document}

\maketitle
\begin{abstract}
What matters more for building foundation models for wireless packet
traces: the tokenizer or the architecture or both? To answer this question,
we build on \ourmethod~\cite{pradhan2026plume}, which
introduced protocol-aware tokenization for 802.11
traces; we scale model depth and transfer the same
tokenizer to a fundamentally different architecture
family.
A deeper GPT (\ourtwo, 24 layers) reaches 98.2\% top-1
accuracy, gaining 32 points over the original 12-layer
design, while a Mamba-2 state-space variant
(\plumemamba) achieves 96.1\% with $1.7\times$ higher
throughput and $2\times$ longer context.
The key insight emerges from a controlled $2{\times}2$
comparison across tokenizers and architectures:
changing the tokenizer swings accuracy by 32 points;
changing the architecture moves it by only 2.
Protocol-aware tokenization is the primary performance
lever, and the backbone becomes a deployment knob
trading accuracy for speed.
\end{abstract}

\section{Introduction}
\label{sec:intro}

Foundation models succeed when their input representation
matches the native structure of the data.
In language, morphology-respecting tokens let models capture
compositional
semantics~\cite{openai_gpt4_2023,chowdhery2022palm};
in vision, pixel-level patches preserve spatial structure.
Wireless packet traces have their own native structure:
layered protocol headers, typed fields, timing gaps,
and cross-packet state machines.
Unlike natural language, where sub-word boundaries carry
morphological cues, protocol field boundaries are
deterministic and carry explicit type information.
A tokenizer that respects this structure compresses
sequences, raises per-token information density, and
provides inductive bias for any downstream architecture.
Ignoring it forces models to spend capacity rediscovering
boundaries that are already present in the protocol
specification.

Prior work introduced
\ourmethod~\cite{pradhan2026plume},
a protocol-aware tokenizer coupled with a GPT backbone
for 802.11 traces.
\ourmethod varied width at fixed depth (12 layers) and
found diminishing returns beyond 225M parameters,
leaving two questions open.
First, \emph{how should the model be scaled?}
We show that depth is the right axis: \ourtwo, a
24-layer GPT, jumps from 66.4\% to 98.2\% top-1 accuracy
(+32 points), the single largest gain in the model family.
Second, \emph{does the tokenizer transfer across
architecture families?}
We pair the same protocol-aware tokenizer with a
Mamba-2~\cite{gu2023mamba} state-space backbone
(\plumemamba), achieving $1.7\times$ higher throughput
and $2\times$ longer context windows than the transformer,
while retaining 96.1\% accuracy.

\begin{figure}[t]
\centering
\resizebox{0.95\columnwidth}{!}{%
\begin{tikzpicture}[
    every node/.style={font=\scriptsize},
    block/.style={draw, rounded corners=2pt, minimum height=0.45cm,
                  align=center, inner sep=2.5pt, font=\scriptsize},
    grp/.style={draw, dashed, rounded corners=3pt, inner sep=5pt},
    annot/.style={font=\tiny\itshape, text=black!55},
    lbl/.style={font=\tiny\bfseries, text=black!65, anchor=east},
    tag/.style={font=\tiny\bfseries, rounded corners=1.5pt,
                inner sep=1.5pt, minimum height=0.3cm},
    arr/.style={->, very thick, draw=black!40,
                >={Stealth[length=5pt, width=4pt]}},
]

\coordinate (labelright) at (0,0);

\node[lbl] at ([xshift=-0.15cm]labelright) (rlbl) {Shared};
\node[block, fill=blue!6, draw=blue!35,
      text width=1.2cm, anchor=west] (pcap) at ([xshift=0.15cm]labelright)
      {802.11\\[-1pt]PCAPs};
\node[block, fill=blue!6, draw=blue!35, right=1.4cm of pcap,
      text width=1.5cm] (dissect) {Wireshark\\[-1pt]Dissector};
\node[block, fill=blue!6, draw=blue!35, right=1.4cm of dissect,
      text width=1.5cm] (tok) {PLUME\\[-1pt]Tokenizer};
\node[block, fill=blue!6, draw=blue!35, right=1.4cm of tok,
      text width=1.2cm] (seq) {Token\\[-1pt]Sequence};
\node[annot, right=0.15cm of seq] (stats) {69K vocab};

\draw[arr] (pcap) -- (dissect);
\draw[arr] (dissect) -- (tok);
\draw[arr] (tok) -- (seq);

\begin{scope}[on background layer]
\node[grp, draw=blue!35, fill=blue!3,
      inner sep=6pt,
      fit=(pcap)(dissect)(tok)(seq)(stats)] (sbox) {};
\end{scope}

\node[lbl] (albl) at ([yshift=-1.6cm]labelright) {Architecture};
\node[block, fill=orange!10, draw=orange!45,
      text width=3.2cm, minimum height=1.0cm,
      anchor=west] (gpt) at ([xshift=0.15cm, yshift=-1.6cm]labelright)
      {\textbf{PLUME-DEEP}\\[1pt]
       24L GPT\;\textbar\;467M\;\textbar\;4K ctx};
\node[annot, right=0.3cm of gpt, yshift=0.1cm] (ga) {98.2\% acc};
\node[annot, below=-0.02cm of ga] (gts) {9,327 tokens/s};
\node[tag, fill=orange!18, draw=orange!45,
      below=0.06cm of gpt] (gtag) {Accuracy-Critical};

\node[block, fill=green!7, draw=green!40!black,
      text width=3.2cm, minimum height=1.0cm,
      anchor=west] (mamba) at ([xshift=1.8cm]ga.east |- gpt.center)
      {\textbf{PLUME-MAMBA}\\[1pt]
       12L Mamba-2\;\textbar\;447M\;\textbar\;8K ctx};
\node[annot, right=0.3cm of mamba, yshift=0.1cm] (ma) {96.1\% acc};
\node[annot, below=-0.02cm of ma] (mts) {16,107 tokens/s};
\node[tag, fill=green!12, draw=green!40!black,
      below=0.06cm of mamba] (mtag) {Speed-Critical};

\begin{scope}[on background layer]
\node[grp, draw=orange!40, fill=orange!3,
      fit=(gpt)(ga)(gts)(gtag)] {};
\node[grp, draw=green!35!black, fill=green!2,
      fit=(mamba)(ma)(mts)(mtag)] {};
\end{scope}

\node[lbl] (olbl) at ([yshift=-3.4cm]labelright) {Downstream};
\node[block, fill=orange!10, draw=orange!45,
      text width=1.6cm, anchor=west] (app1)
      at ([xshift=0.15cm, yshift=-3.4cm]labelright)
      {Anomaly\\[-1pt]Detection};
\node[block, fill=orange!10, draw=orange!45,
      right=1.5cm of app1, text width=1.6cm] (app2)
      {Root-Cause\\[-1pt]Analysis};
\node[block, fill=gray!10, draw=gray!45,
      right=1.5cm of app2, text width=1.6cm] (app3)
      {Traffic\\[-1pt]Forecasting};
\node[block, fill=green!7, draw=green!40!black,
      right=1.5cm of app3, text width=1.6cm] (app4)
      {Real-Time\\[-1pt]Monitoring};

\end{tikzpicture}%
}
\caption{System overview.
802.11 PCAPs are encoded via a protocol-aware
tokenizer, then modeled by PLUME-DEEP
(accuracy-critical, \textcolor{orange!70!black}{orange}) or
\plumemamba (speed-critical, \textcolor{green!50!black}{green}).
Downstream task colors indicate architecture suitability;
grey denotes tasks where either may apply.}
\label{fig:overview}
\end{figure}
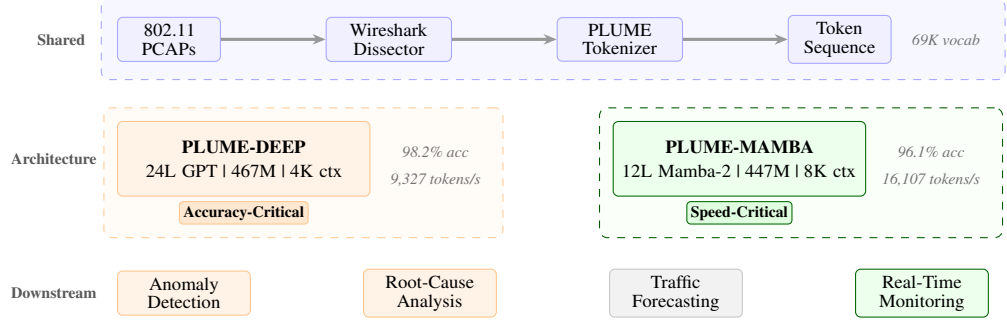

Figure~\ref{fig:overview} summarizes the pipeline.
To disentangle the effects of tokenization and
architecture, we construct a full $2{\times}2$ comparison
(Table~\ref{tab:full_results}).
\netssm~\cite{chu2026netssm} supplies the
protocol-agnostic Mamba cell: it uses the same architecture
as \plumemamba but a generic byte-level tokenizer.
We additionally train a byte-level tokenized GPT
(Byte-GPT) to fill the remaining cell.
The results are striking.
Swapping the tokenizer on the same Mamba backbone produces
a 32-point accuracy swing (96.1\% vs.\ 64.1\%); swapping
the architecture under the same protocol-aware tokenizer
moves accuracy by only 2 points (98.2\% vs.\ 96.1\%).
Tokenization is the primary performance lever.

Three key findings are the following:

\begin{enumerate}[leftmargin=*]
\item \textbf{Depth is the scaling axis for protocol
  sequences.}
  Going from 12 to 24 layers at comparable parameter count
  produces +32 points of accuracy and a $12\times$ loss
  reduction.
  Width scaling and longer context windows both hurt at
  this data scale.

\item \textbf{Mamba is faster with longer context;
  the transformer is more accurate.}
  \plumemamba delivers $1.7\times$ higher throughput
  (16,107 vs.\ 9,327\,tok/s) and $2\times$ context
  (8K vs.\ 4K tokens), while \ourtwo leads on
  accuracy (98.2\% vs.\ 96.1\%) and loss (33\% lower).
  The right choice depends on the deployment:
  accuracy-critical tasks favor the transformer,
  latency-sensitive monitoring favors Mamba,
  and protocols with long transactions benefit from the
  larger context.

\item \textbf{Tokenization is the primary lever;
  architecture is a deployment knob.}
  Both protocol-aware models exceed 96\% accuracy; both
  generic-tokenizer baselines remain below 65\%.
  On the Mamba backbone alone, replacing generic tokens with
  protocol-aware tokens accounts for a 32-point gain,
  while on the GPT side the gap is even wider (51 points).
  Protocol-aware representation matters more than
  architecture choice for structured packet data.
\end{enumerate}

\subsection{Contributions}

\begin{enumerate}[leftmargin=*]
\item \textbf{\ourtwo}, a 467M-parameter, 24-layer
  GPT that achieves 98.2\% top-1 accuracy and 1.10
  perplexity on 802.11 next-packet prediction
  (Table~\ref{tab:full_results}).

\item \textbf{\plumemamba}, a Mamba-2 SSM that reuses the
  protocol-aware tokenizer, reaching 96.1\% top-1 at
  $1.7\times$ higher throughput and $2\times$ longer context
  than \ourtwo (\S\ref{sec:exp:head_to_head}).

\item A controlled $2{\times}2$ comparison across
  tokenizer and architecture axes, establishing that
  protocol-aware tokenization is the dominant factor and
  that depth outperforms width for structured protocol
  data (Tables~\ref{tab:full_results},~\ref{tab:gpt_scaling}).

\end{enumerate}

\section{Background}
\label{sec:background}

\subsection{Protocol-Aware Tokenization}
\label{sec:background:tokenization}

Wireless packet traces are structured data.
Each packet is a hierarchy of protocol layers
(802.11, IP, TCP, EAPOL, DNS, etc.),
each layer contains typed fields
(addresses, flags, counters, timestamps),
and packets form conversations governed by state machines
(authentication handshakes, association sequences,
key exchanges).
Standard tokenizers ignore all of this.

\textbf{Why generic tokenizers fail.}
Generic tokenizers such as BPE~\cite{sennrich2016bpe}
and byte-level encodings~\cite{chu2026netssm} treat input as flat text,
seeking reusable sub-word or byte units optimized for
natural language.
Applied to protocol dissections, they fragment field
boundaries: a single field such as
\texttt{wlan.fc.type\_subtype} becomes many byte tokens,
each carrying little protocol meaning.
The result is a $6.2\times$ sequence inflation relative
to field-level tokenization and a drop in per-token
entropy from 7.61 to 6.70 bits~\cite{pradhan2026plume}.

\textbf{Protocol-aware tokenization.}
We use the tokenizer from
\ourmethod~\cite{pradhan2026plume},
which makes four design choices:
(1)~one token per dissector field name
(e.g., \texttt{wlan.fc.type\_subtype} is atomic);
(2)~type-aware value encoding (symbols expanded to words,
numerics preserved, strings sub-tokenized);
(3)~explicit layer boundary markers; and
(4)~gap tokens for inter-arrival times.
The vocabulary contains 69,842 tokens.
A typical packet is 327 tokens (vs.\ 2,014 for byte-level),
so a 4,096-token context window holds ${\sim}$12 packets,
enough for a complete 802.11 authentication exchange.
The $6.2\times$ compression ratio means that for the same
context budget, protocol-aware tokenization covers
substantially more packets, giving the model a wider view
of the conversation state without increasing computational
cost.

\subsection{PLUME Primer}
\label{sec:background:plume}

\ourmethod~\cite{pradhan2026plume} (Protocol Language
Understanding Model for Exchanges) is the foundation that
this paper extends.

\textbf{Pipeline.}
Raw PCAP files are decoded with Wireshark's dissector
engine into a tree of protocol layers and typed fields.
The protocol-aware tokenizer flattens this tree into a
token sequence preserving field boundaries and layer
hierarchy, which is fed to a causal language model
trained with next-token prediction (cross-entropy loss).

\textbf{Architecture.}
The original \ourmethod is a 12-layer GPT with
768-dimensional embeddings, 12 heads, a 2,048-token
context window, and 140M parameters, trained on the
same corpus described below.
A key insight from \ourmethod is that the tokenizer
and model are co-designed: the protocol-aware vocabulary
gives the GPT backbone a structured input where
each token carries field-level semantics,
removing the need for the model to discover protocol
boundaries from raw bytes.

\textbf{Anomaly detection via loss.}
At inference, per-token cross-entropy loss scores each
packet: normal packets produce low loss; anomalous
packets spike because unexpected field values deviate
from learned patterns.
Thresholding per-packet average loss enables zero-shot
anomaly detection, and each high-loss token directly
identifies the anomalous protocol field.
This loss-based approach requires no labeled anomaly
data and generalizes to unseen failure modes, making it
practical for enterprise deployments where new fault
types emerge continuously.

\textbf{Key results and open questions.}
\ourmethod achieved 83.1\% token accuracy and effective
anomaly detection across five failure categories.
Width scaling (768$\to$1536 embedding at 12 layers)
yielded diminishing returns due to overfitting,
motivating the depth-scaling investigation in this paper.
\ourmethod also left open whether its tokenizer is
architecture-specific or transferable; we answer this
question by pairing it with a Mamba-2 SSM backbone.

\subsection{Training Data and Evaluation}
\label{sec:background:data}

All models in this paper train on the same curated corpus:
7,890 PCAP files (149K packets, 48.9M tokens) from
enterprise 802.11 deployments, with beacon dominance
reduced from ${>}$50\% to 4.7\% via
HDBSCAN~\cite{mcinnes2017hdbscan}
and MMR~\cite{carbonell1998mmr}
sampling.
We evaluate on five real-world failure categories
(Bad Password, EAPOL Timeout, Invalid PMKID,
Unable to Handle New STA, Rejected Temporarily),
50 PCAPs each.
These categories were selected because they represent
the most frequent root-cause classes in enterprise
wireless troubleshooting and span different protocol
stages (authentication, association, key exchange),
testing whether the model captures state-machine logic
across the 802.11 connection lifecycle.

\section{Method}
\label{sec:method}

We investigate two axes of improvement over the original
\ourmethod~\cite{pradhan2026plume} GPT backbone:
scaling depth (\S\ref{sec:method:depth}) and replacing
the transformer with a Mamba-2 state-space model
(\S\ref{sec:method:mamba}).
Both variants share the same protocol-aware tokenizer; the only variables are architecture and
depth.

\subsection{Scaling Depth: \ourtwo}
\label{sec:method:depth}

The original \ourmethod width-scaling study
(12 layers, 768$\to$1536 embedding) showed diminishing
returns beyond 225M parameters: the 450M wide model
overfit severely.
Under our evaluation protocol (\S\ref{sec:exp:depth}),
the same 12-layer, 1536-dimensional configuration reaches
only 66.4\% top-1 accuracy, confirming that width alone
does not help at this data scale.
Our hypothesis is that for protocol sequences, where
meaning is hierarchical (layer, field, value) and
dependencies span multiple packets, depth provides a
stronger inductive bias than width.

\ourtwo doubles transformer depth to 24 layers while
reducing the embedding dimension to 1152 (18 heads),
yielding 467M parameters at a 4,096-token context window.
The narrower but deeper design keeps the parameter budget
comparable to the 450M wide model while allocating
capacity to compositional feature extraction across
layers.
Each additional layer provides a new level of abstraction:
early layers encode individual field values, middle layers
compose fields into packet-level representations, and the
deepest layers capture cross-packet dependencies such as
handshake state transitions. Training uses AdamW~\cite{loshchilov2019adamw}
($\beta_1{=}0.9$, $\beta_2{=}0.95$),
learning rate $7{\times}10^{-4}$ with cosine decay,
100 warmup iterations, gradient clipping at 1.0,
2,000 iterations with effective batch size 12.
All models are trained from scratch (no pre-trained
initialization).

\textbf{Data-to-parameter ratio.}
The training corpus contains 48.9M tokens
(149K packets).
At 467M parameters, \ourtwo operates at roughly 0.1
tokens per parameter, well below the compute-optimal
ratios suggested by
Chinchilla~\cite{hoffmann2022chinchilla}.
This data-constrained regime is representative of real
enterprise deployments where labeled wireless captures
are scarce.
The depth-scaling result should be interpreted within this
context: the narrow-deep architecture (24L, 1152d) avoids
overfitting because each layer has fewer parameters than
the wide-shallow alternative (12L, 1536d), reducing
per-layer memorization capacity.
Whether the depth advantage persists at 10 to
100$\times$ more data is an open question for future work.

\subsection{\plumemamba: Protocol Tokens with SSM Backbone}
\label{sec:method:mamba}

Transformers have quadratic attention complexity in sequence
length, limiting throughput for long contexts.
Mamba-2~\cite{gu2023mamba} replaces attention with selective
state-space layers that process sequences in linear time,
enabling longer context windows without throughput loss.

\plumemamba pairs the protocol-aware tokenizer from
\ourmethod with a Mamba-2 backbone:
12 layers, 1536 embedding dimension, expand factor 4,
state dimension ($d_\text{state}$) 128,
8,192-token context, 447M parameters.
This is the same architecture as \netssm but with a
different tokenizer, enabling a controlled comparison
of tokenizer effect on the SSM backbone. Training uses the same optimizer as the GPT models:
AdamW ($\beta_1{=}0.9$, $\beta_2{=}0.95$),
learning rate $5{\times}10^{-4}$ with cosine decay,
100 warmup iterations, gradient clipping at 1.0,
2,000 iterations with effective batch size 12.
The lower learning rate (vs.\ $7{\times}10^{-4}$ for GPT)
was selected via preliminary sweeps to stabilize SSM
training.

Linear-time complexity gives \plumemamba two practical
advantages over \ourtwo:
(1)~$1.7\times$ higher throughput
(16,107 vs.\ 9,327\,tok/s), and
(2)~$2\times$ longer context (8,192 vs.\ 4,096
tokens, covering roughly 25 packets instead of 12).
The accuracy cost is modest: 96.1\% vs.\ 98.2\% top-1,
placing both protocol-aware models in the same
performance tier and far above generic-tokenizer
baselines.
The gap arises because Mamba's selective state-space
layers compress the history into a fixed-dimensional
state vector, which can discard rare field-value
co-occurrences that quadratic attention preserves by
attending to every prior token explicitly.

\subsection{Baseline: \netssm}
\label{sec:method:netssm}

\netssm~\cite{chu2026netssm} is a Mamba-2 state-space
model designed for network traffic generation.
It uses 24 layers with a 768-dimensional embedding,
expand factor 4, a 4,096-token context, and 170M
parameters.
Its tokenizer does not enforce protocol-layer hierarchy
or field-boundary alignment.
We include \netssm to isolate the tokenizer's
contribution: \plumemamba and \netssm share the same
architecture family but differ only in tokenization,
creating a natural ablation.

\section{Experiments}
\label{sec:experiments}

We organize evaluation around four questions:
(1)~how do the architectures compare overall
(\S\ref{sec:exp:results})?
(2)~how does depth scaling compare to width and context
scaling within GPT (\S\ref{sec:exp:depth})?
(3)~what are the specific trade-offs between \ourtwo
and \plumemamba (\S\ref{sec:exp:head_to_head})?
(4)~does the tokenizer transfer across architectures
(\S\ref{sec:exp:tokenizer})?

\subsection{Setup}
\label{sec:exp:setup}

\textbf{Models.}
Table~\ref{tab:full_results} lists all configurations.
All \ourmethod variants share the same 69K-token vocabulary
and curated training data.
\netssm uses its own tokenizer and pipeline.
\plumemamba and \netssm share the same Mamba architecture
but differ in tokenization, isolating the tokenizer effect.

\textbf{Hardware.}
All models train and evaluate on a single NVIDIA A10G
(24\,GB GDDR6) on an AWS \texttt{g5.12xlarge} instance.

\textbf{Throughput measurement.}
Throughput (tok/s) is measured during autoregressive
\emph{generation} (not prefill): we generate next-packet
sequences token by token with batch size 1, and report
tokens produced per wall-clock second averaged over the
full test set (250 PCAPs).
This reflects end-to-end inference cost for deployment.

\textbf{Metrics.}
We report five metrics to capture complementary aspects of
model quality.
\emph{Top-1 accuracy} measures whether the highest-probability
token matches the ground truth; \emph{top-5 accuracy} relaxes
this to the top five predictions, reflecting how often the
correct token is among plausible candidates.
\emph{Cross-entropy loss} quantifies the model's uncertainty
over the full vocabulary at each position; lower loss
indicates sharper, more confident predictions.
\emph{Perplexity} (the exponentiated loss) gives an intuitive
reading: a perplexity of 1.10 means the model is, on average,
choosing among roughly 1.1 equally likely tokens.
\emph{Throughput} (tokens per second) measures end-to-end
inference speed under autoregressive generation.

\textbf{Test data.}
We evaluate on five categories of real-world 802.11
authentication failures sourced from enterprise deployments,
with 50 packet captures (PCAPs) per category (250 total).
The categories are:
(1)~\emph{Bad Password}, where the client supplies incorrect
credentials during the EAP exchange;
(2)~\emph{EAPOL Timeout}, where the four-way handshake fails
to complete within the expected window;
(3)~\emph{Invalid PMKID}, where the pairwise master key
identifier does not match the access point's expectation;
(4)~\emph{Unable to Handle New STA}, where the AP rejects a
new station due to resource constraints; and
(5)~\emph{Rejected Temporarily}, where the AP defers
association with a temporary status code.
These categories cover distinct failure modes spanning
credential errors, timing issues, key-management faults,
and capacity limits, providing a diverse evaluation surface.

\subsection{Results}
\label{sec:exp:results}

\begin{table}[t]
\centering
\caption{Full results across architectures and tokenizers.
All \ourmethod variants share the same protocol-aware
tokenizer (vocab $\approx$69K). \netssm uses its own
tokenizer. Best per metric result is in \textbf{bold}.
\ourmethod~1.0 accuracy is token accuracy reported in the
original paper (different evaluation protocol).}
\label{tab:full_results}
\small
\begin{tabular}{llccccccc}
\toprule
Model & Tokenizer & Params & Ctx
  & Top-1 & Top-5 & Loss & PPL & Tok/s \\
\midrule
\ourmethod 1.0 & Protocol & 140M & 2K
  & 83.1\% & --- & 0.42 & 1.52 & --- \\
\ourtwo & Protocol & 467M & 4K
  & \textbf{98.19\%} & \textbf{99.26\%}
  & \textbf{0.092} & \textbf{1.10} & 9,327 \\
\plumemamba & Protocol & 447M & 8K
  & 96.14\% & 99.01\% & 0.140
  & 1.15 & 16,107 \\
\netssm & Generic & 170M & 4K
  & 64.09\% & 75.19\% & 1.66
  & 5.27 & \textbf{35,579} \\
Byte-GPT & Generic &346M & 4K & 47.57\%
& 61.59\% & 2.69 & 14.76 & 17,477 \\
\bottomrule
\end{tabular}
\end{table}

\begin{table}[h]
\centering
\caption{Macro-averaged performance across the five error
categories (Bad Password, EAPOL Timeout, Invalid PMKID,
Unable to Handle New STA, Rejected Temporarily).
Best per metric result is in \textbf{bold}.}
\label{tab:avg_results}
\small
\begin{tabular}{lrrrrrr}
\toprule
Model & Params & PPL & Loss & Top-1 & Top-5 & Tok/s \\
\midrule
\ourtwo      & 467M & 1.64 & 0.492
  & \textbf{95.3\%} & 96.7\% & 11,012 \\
\plumemamba  & 447M & \textbf{1.50} & \textbf{0.405}
  & 94.6\% & \textbf{97.2\%} & 16,008 \\
\netssm      & 170M & 5.15 & 1.610
  & 67.2\% & 76.0\% & \textbf{35,701} \\
\bottomrule
\end{tabular}
\end{table}

Table~\ref{tab:avg_results} reports macro-averaged
metrics across the five failure categories.
The category-level view reveals a nuance not visible in
the aggregate: \plumemamba achieves lower perplexity and
loss than \ourtwo (1.50 vs.\ 1.64 PPL), even though
\ourtwo leads on top-1 accuracy (95.3\% vs.\ 94.6\%).
This suggests that the SSM backbone produces a more
calibrated distribution across categories, while the
transformer concentrates its advantage on the most
common token patterns.
The gap between protocol-aware models and \netssm remains
large (roughly 28 points in top-1), reinforcing the
tokenizer effect observed in the aggregate results.

\ourtwo achieves the highest accuracy
(98.2\% top-1) and lowest loss (0.092), while
\plumemamba leads on throughput (16,107\,tok/s), reflecting
linear-time SSM complexity.
The gap between \plumemamba and \netssm is the most
informative row in the table: both use the same Mamba
backbone, yet protocol-aware tokenization raises
top-1 accuracy from 64.1\% to 96.1\%, a 32-point
improvement that dwarfs the 2-point difference between
\ourtwo and \plumemamba.
\ourmethod~1.0 (12L, 140M) reaches 83.1\% under its
original evaluation protocol; depth scaling to \ourtwo
provides the dominant improvement axis.

\subsection{GPT Depth Scaling}
\label{sec:exp:depth}

Table~\ref{tab:gpt_scaling} isolates depth vs.\ width vs.\
context length within the GPT family.

\begin{table}[t]
\centering
\caption{GPT depth scaling. All use the same tokenizer and
training data. Depth is the only axis that improves
accuracy at this data scale.}
\label{tab:gpt_scaling}
\small
\begin{tabular}{lcccccr}
\toprule
Config & Layers & Embd & Context
  & Top-1 & PPL & Loss \\
\midrule
PLUME 1.0 & 12 & 768 & 2,048
  & 83.1\% & 1.52 & 0.420 \\
Shallow @ 8K & 12 & 1536 & 8,192
  & 54.5\% & 6.65 & 1.894 \\
Shallow @ 4K & 12 & 1536 & 4,096
  & 66.4\% & 3.05 & 1.116 \\
Deep @ 4K & 24 & 1152 & 4,096
  & \textbf{98.2\%} & \textbf{1.10}
  & \textbf{0.092} \\
\bottomrule
\end{tabular}
\end{table}

\textbf{Depth is key.}
Going from 12 to 24 layers at comparable parameter count
produces +32 points of accuracy
(66.4\%$\to$98.2\%) and $12\times$ loss reduction
(1.116$\to$0.092).

\textbf{Context length hurts at fixed depth.}
Doubling the context window from 4K to 8K at fixed depth
\emph{reduces} accuracy (66.4\%$\to$54.5\%).
802.11 prediction accuracy saturates at two to three
context packets~\cite{pradhan2026plume}; longer windows
dilute attention over irrelevant distant tokens.

\textbf{Why depth helps.}
Protocol sequences are hierarchical: layer boundaries
nest fields, fields carry typed values, cross-packet
dependencies span authentication handshakes.
At 12 layers, the model captures local field
co-occurrences but misses multi-packet state-machine
transitions.
At 24 layers, it models the full handshake grammar.
This aligns with findings in NLP, where deeper
transformers learn increasingly abstract syntactic and
semantic features at successive layers; in our setting,
the ``syntax'' is the protocol state machine and the
``semantics'' are the field-value dependencies that
determine whether a connection succeeds or fails.

\subsection{Head-to-Head: \ourtwo vs.\ \plumemamba}
\label{sec:exp:head_to_head}

This comparison uses each architecture at its
\emph{best available configuration}: the deepest GPT
(24L, 4K context) against the Mamba model (12L, 8K
context).
Because depth and context length differ between the two
models, the accuracy gap cannot be attributed to
architecture alone; Table~\ref{tab:gpt_scaling} shows
that both variables impact performance.
A matched-depth, matched-context comparison (12L, 4K for
both) is planned to isolate the pure architectural effect.
The value of this comparison is \emph{practical}: it
shows the best achievable operating point for each
architecture family under our training budget.


\ourtwo leads on every accuracy metric:
98.2\% vs.\ 96.1\% top-1, 99.3\% vs.\ 99.0\% top-5,
and perplexity 1.10 vs.\ 1.15.
The 33\% loss reduction means \ourtwo assigns higher
probability to correct next tokens, which translates to
sharper anomaly signals (lower false-positive rates at
fixed thresholds) and more reliable generation.
Concretely, when \ourtwo encounters an unexpected field
value, such as a malformed PMKID or an out-of-sequence
EAPOL frame, the loss spike is sharper and easier to
localize, allowing detection thresholds to be set without
per-category tuning.

\begin{figure}[t]
  \centering
  \begin{subfigure}[t]{0.45\columnwidth}
    \centering
    \includegraphics[width=\linewidth]{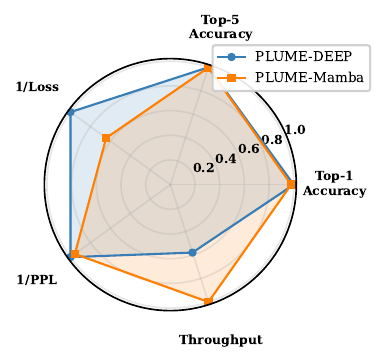}
    \caption{Head-to-head radar. \ourtwo dominates accuracy;
    \plumemamba dominates throughput.}
    \label{fig:radar}
  \end{subfigure}%
  \hfill
  \begin{subfigure}[t]{0.45\columnwidth}
    \centering
    \includegraphics[width=\linewidth]{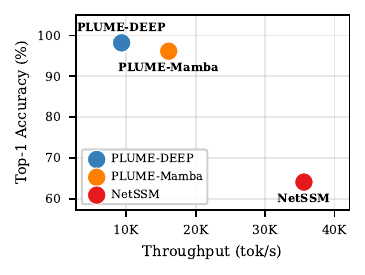}
    \caption{Speed--accuracy Pareto frontier. Deployment
    scenario determines the operating point.}
    \label{fig:pareto}
  \end{subfigure}
  \caption{Head-to-head comparison of \ourtwo (transformer,
  24L, 4K context) and \plumemamba (Mamba-2 SSM, 12L, 8K
  context).\ourtwo leads on accuracy; \plumemamba leads on
  throughput and context length, defining complementary
  operating points.}
  \label{fig:tradeoffs}
\end{figure}

\plumemamba wins on throughput ($1.7\times$) and context
length ($2\times$), as visualized in
Figure~\ref{fig:tradeoffs}.
The throughput gain follows directly from linear-time
recurrence versus quadratic attention.
The longer context window is valuable for protocols with
extended transaction sequences, such as enterprise RADIUS
authentication chains that span dozens of packets.

\subsection{Tokenizer Effect: \plumemamba vs.\ \netssm}
\label{sec:exp:tokenizer}

\plumemamba and \netssm share the same Mamba-2 architecture
(12 layers, 1536 embedding, 447M parameters, 8K context).
The only difference is the tokenizer: protocol-aware
(69K vocabulary, ${\sim}$327 tokens/packet) vs.\ generic
byte-level (${\sim}$2,014 tokens/packet).

\textbf{Results.}
Table~\ref{tab:full_results} shows that \plumemamba
outperforms \netssm on every quality metric.
Top-1 accuracy rises from 64.1\% to 96.1\%, top-5
accuracy from 75.2\% to 99.0\%, loss drops from 1.66
to 0.14, and perplexity falls from 5.27 to 1.15.
The only trade-off is throughput: \plumemamba's
larger vocabulary produces longer token IDs, reducing
throughput relative to \netssm's compact byte-level tokens.
Because the architecture is held constant, this
32-point accuracy gain is attributable entirely to the
tokenizer, providing the strongest evidence in our
study that tokenizer choice is the primary driver of
model quality.

\textbf{Structural advantages of protocol-aware tokens.}
Independent of aggregate accuracy, protocol-aware
tokenization provides two benefits:
(1)~$6.2\times$ sequence compression means each token
carries more semantic weight, improving interpretability
and per-token anomaly localization (a high-loss token
directly identifies which protocol field is anomalous);
and (2)~field-aligned boundaries enable structured
generation of valid protocol messages without
post-hoc parsing.

\section{Analysis}
\label{sec:analysis}

\subsection{Why Protocol-Aware Tokenization Is the Primary Lever}
\label{sec:analysis:tokenizer}

Across two architecture families, protocol-aware
tokenization consistently enables strong performance.
The reason is that the tokenizer encodes domain knowledge
that the model would otherwise have to learn from data
alone.

\textbf{Tokenizer transfers across architectures.}
\ourtwo (98.2\%) and \plumemamba (96.1\%) both use
protocol-aware tokens on different backbones, and both
outperform \ourmethod~1.0 (83.1\%) and \netssm (64.1\%).
No architecture-specific tuning was required, suggesting
that the inductive bias the tokenizer provides is
orthogonal to the choice of sequence model.

\begin{figure}[t]
  \centering
  \includegraphics[width=0.60\columnwidth]{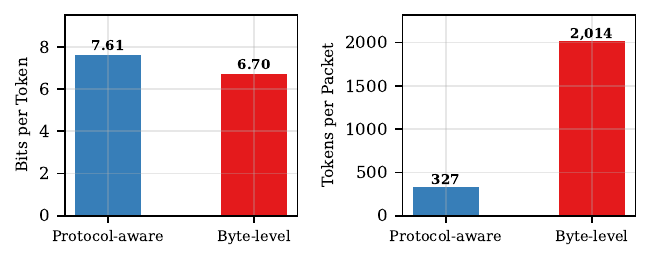}
  \caption{Per-token information density. Protocol-aware
  tokens carry 7.61 bits vs.\ 6.70 for byte-level.}
  \label{fig:tokenizer_effect}
\end{figure}

\textbf{Compression drives per-token signal.}
Protocol-aware tokenization compresses packets by
$6.2\times$ relative to byte-level tokenization
(Figure~\ref{fig:tokenizer_effect}).
Each token maps to a semantically meaningful unit: a
field name, a typed value, or a layer boundary.
This compression is not merely a sequence-length
reduction; it concentrates information.
Per-token entropy rises from 6.70 to 7.61 bits, meaning
each prediction the model makes carries more semantic
weight.
For anomaly detection, the practical consequence is that
a high-loss token directly identifies which protocol
field is unexpected.

\textbf{The tokenizer shapes the learning problem.}
Without field-boundary alignment, a model must learn
co-occurrence patterns across five to ten byte tokens to
reconstruct a single protocol field.
With protocol-aware tokens, the model predicts field
values conditioned on field names and neighbors, a
simpler conditional distribution.
This explains why \netssm scores 32 points below
\plumemamba as it spends
capacity on segmenting fields rather than learning
protocol semantics.

\subsection{Depth vs.\ Width for Structured Data}
\label{sec:analysis:depth_width}

The original \ourmethod width-scaling
study~\cite{pradhan2026plume} showed that widening from
768 to 1536 at 12 layers caused overfitting; the same
configuration reaches only 66.4\% under our evaluation
protocol (Table~\ref{tab:gpt_scaling}).
\ourtwo reverses this trend: deepening from 12 to 24
layers produces a +32-point gain at comparable parameter
count.

The explanation lies in the hierarchical nature of
protocol data.
Each transformer layer composes features at one level
of abstraction.
At 12 layers, the model learns field-level patterns
(address prediction, frame-type classification) but
misses multi-step protocol logic; the four-way EAPOL
handshake, for example, requires tracking state across
four or more packets, each containing 300+ tokens.
At 24 layers, the model composes packet-level features
into conversation-level representations.
The wide model fails not for lack of parameters but
because 12 layers cannot compose these hierarchical
features, and the excess per-layer capacity memorizes
training-specific patterns.

This result holds in a data-constrained regime (roughly
0.1 tokens per parameter); at Chinchilla-optimal data
scales the depth advantage may shrink.
Nonetheless, for enterprise deployments where labeled
captures are scarce, depth scaling is the more robust
strategy, and the principle likely extends to other
hierarchical domains (genomics, chemistry, industrial
control) where the vocabulary already encodes domain
structure.

\subsection{When to Use Which Architecture}
\label{sec:analysis:tradeoff}

Because both protocol-aware models exceed 96\% accuracy,
the choice of backbone reduces to deployment constraints
rather than fundamental quality differences.

\textbf{Accuracy-critical: \ourtwo.}
At 98.2\% top-1 and 1.10 perplexity, \ourtwo suits
offline root-cause analysis and forensic investigation,
where a loss spike pinpoints the exact field that
deviates from the expected handshake.

\textbf{Speed-critical: \plumemamba.}
With $1.7\times$ higher throughput and 96.1\% accuracy,
\plumemamba suits real-time monitoring of high-volume
access points where thousands of packets per second must
be scored.
The 2-point accuracy cost is acceptable when the goal is
stream-level anomaly flagging rather than per-packet
forensics.
In practice, the linear-time complexity of Mamba means
that throughput scales gracefully as packet rates
increase, whereas the transformer's quadratic attention
becomes the bottleneck.

\textbf{Long-context: \plumemamba.}
The $2\times$ larger context window
(8K tokens, roughly 25 packets) makes \plumemamba
valuable for protocols with extended multi-step
transactions, including enterprise RADIUS chains,
multi-round EAP exchanges, and roaming sequences that
span many packets.
\ourtwo's 4K context (roughly 12 packets) suffices for
standard 802.11 authentication but falls short for longer
conversations.
This distinction becomes increasingly important as
enterprise networks adopt more complex authentication
schemes that involve longer packet exchanges.

The practical guideline is: invest in the tokenizer
first, then select the architecture based on the
deployment profile.

\section{Related Work}
\label{sec:related}

\textbf{Foundation models for network traffic.}
Lens~\cite{wang2024lens} pre-trains via masked span
prediction for traffic classification.
netFound~\cite{guthula2023netfound} pre-trains on
unlabeled traces with self-supervised embeddings
(640M parameters).
NetGPT~\cite{meng2023netgpt} tokenizes multi-pattern
traffic into unified text, and
LLMcap~\cite{tulczyjew2024llmcap} applies masked
language modeling to PCAPs.
All of these systems use generic tokenization and
primarily target wired or encrypted traffic.
TrafficGPT~\cite{qu2024trafficgpt} extends the token
window to 12K via linear attention and reversible
tokenization, but does not exploit protocol structure.
Our results show that protocol-aware tokenization lifts
both transformer and SSM architectures on 802.11 data,
a complementary finding.

\textbf{State-space models for sequences.}
Mamba~\cite{gu2023mamba} introduced selective SSMs with
linear-time complexity.
NetMamba~\cite{wang2024netmamba} applies Mamba to
encrypted traffic classification, while
\netssm~\cite{chu2026netssm} adapts Mamba-2 for network
traffic generation with a byte-level tokenizer.
By pairing the same Mamba architecture with protocol-aware
tokens (\plumemamba), we isolate the tokenizer's
contribution and show that the structural benefits of
protocol-aware tokenization (compression,
interpretability) complement any architecture choice.

\textbf{Scaling laws.}
Chinchilla~\cite{hoffmann2022chinchilla} established
compute-optimal scaling ratios, and
Kaplan et al.~\cite{kaplan2020scalinglawsneurallanguage}
demonstrated power-law scaling with model size.
Our depth-scaling results complement these general laws
with a domain-specific finding: for structured protocol
data in a data-constrained regime, depth outperforms
width.

\textbf{Domain-specific tokenization.}
The Byte Latent
Transformer~\cite{pagnoni2024blt} shows that dynamic
patching can rival fixed vocabularies, and
DBF-PSR~\cite{Ding2025DBF-PSR} uses protocol semantic
representations.
Our work extends this line of evidence by showing that a
domain-native tokenizer transfers across architecture
families and is the primary performance lever.

\section{Conclusion}
\label{sec:conclusion}

The central finding of this work is that for structured
packet data, the tokenizer matters more than the
architecture.
A $2{\times}2$ comparison across tokenizers and
backbones shows a 32-point accuracy swing from the
tokenizer axis versus a 2-point swing from the
architecture axis.
Once a protocol-aware tokenizer is in place, both a
24-layer transformer (\ourtwo, 98.2\% accuracy) and a
Mamba-2 SSM (\plumemamba, 96.1\% accuracy) operate in
the same high-performance tier, and the backbone choice
reduces to deployment constraints: accuracy versus
throughput, short versus long context. A secondary finding is that depth, not width, is the
right scaling axis for protocol sequences.
Doubling transformer depth produces a 32-point accuracy
gain, while widening at fixed depth yields diminishing
returns and eventual overfitting in this domain.

\textbf{Limitations and future work.}
Our evaluation covers a single enterprise 802.11
deployment; multi-site and cross-protocol generalization
remain open.
The cross-tokenizer comparison is limited by the fact
that token-level metrics are not directly comparable
across vocabularies of different granularity;
tokenizer-agnostic metrics such as bits-per-byte and
packet-level accuracy are needed for a definitive
ablation.
The architectural comparison is confounded by differing
depth and context length; a matched-configuration
experiment is planned.
All results are single-seed; multi-run confidence
intervals will strengthen the conclusions.
Extending protocol-aware tokenization to wired protocols
(TCP/IP, QUIC) and industrial control systems (Modbus,
OPC-UA) is a natural next step.

\textbf{Broader perspective.}
The \ourmethod family illustrates a general principle:
when data has deterministic structure (typed fields,
hierarchical nesting, state-machine semantics), a
tokenizer that respects that structure provides
inductive bias that any downstream architecture
can exploit.
Architecture matters, but representation is the
foundation.
We believe this principle extends to genomics,
chemistry, and industrial control, and that the
co-design template demonstrated here applies wherever
domain-native tokenization is feasible.

\bibliographystyle{plainnat}
\bibliography{references}

\end{document}